\documentclass[aps,prb,amsmath,twocolumn,superscriptaddress,floatfix,footinbib,longbibliography]{revtex4}

\usepackage{graphicx}
\usepackage{epstopdf}

\usepackage[T1]{fontenc}
\usepackage[applemac]{inputenc}
\usepackage{lmodern}
\usepackage[english]{babel}
\usepackage{ae}
\usepackage{units}

\usepackage{amsmath,amssymb,natbib,bm}
\usepackage{psfrag}
\usepackage{subfigure}
\usepackage{amsthm}

\usepackage{fixmath}
\usepackage{booktabs}

\usepackage{slashed}

\usepackage[americaninductors]{circuitikz}
\usepackage{tikz}
\usetikzlibrary{arrows}
\usepackage{url}

\usepackage[colorlinks]{hyperref}

\usepackage{tabularx}

\hypersetup{%
        plainpages=true,
        breaklinks=true,
        hypertexnames=false,
        pageanchor=true,
        colorlinks=true,
        linkcolor={blue},
        citecolor={magenta},
        urlcolor={blue},
        anchorcolor={black}
      }

\renewcommand{\eqref}[1]{\mbox{Eq.~(\ref{#1})}}

\newcommand{\abs}[1]{\left|#1\right|}

\newcommand{\be}{\begin{equation}}
\newcommand{\ee}{\end{equation}}
\newcommand{\bea}{\begin{eqnarray}}
\newcommand{\eea}{\end{eqnarray}}

\begin{document}

\title{Interferometry and dynamics of a transmon-type qubit in front of a mirror}

\author{M.~P.~Liul}
\email[e-mail: ]{liul@ilt.kharkov.ua}
\affiliation{B.~Verkin Institute for Low Temperature Physics and Engineering, Kharkov 61103, Ukraine}

\date{\today}
\begin{abstract}
We theoretically describe the stationary regime and coherent dynamics of a capacitively shunted transmon-type qubit which is placed in front of a mirror. The considered qubit is irradiated by two signals: pump (dressing) and probe. By changing amplitudes and frequencies of these signals we study the system behaviour. The main tool of our theoretical analysis is solving of the Lindblad equation. We also consider Lindblad superoperators in charge and energy bases and compare the results. Theoretically obtained occupation probability is related to the experimentally measured value. This study helps to understand better the properties of qubit-mirror system and gives new insights about the underlying physical processes.

\end{abstract}

\maketitle

\section{Introduction}
Considerable attention is being drawn to topics related to quantum computing [\onlinecite{Nielsen2010}]. Among the promising building blocks for such devices, superconducting qubits stand out, operating at nanosecond scales with millisecond coherence times [\onlinecite{Kockum2019}]. These qubits, controlled by microwaves and featuring lithographic scalability [\onlinecite{Oliver2013}], hold great potential for the advancement of quantum computers.

An essential aspect is the exploration of superconducting qubits within a semi-infinite transmission line [\onlinecite{Hoi2015}], crucial for quantum electrodynamics, particularly waveguide quantum electrodynamics (WQED) [\onlinecite{Kannan2020}]. Notably, in Ref.~[\onlinecite{wen18}], it was revealed that a transmon qubit at the transmission line's end could amplify a probe signal with an amplitude gain of up to 7\%. In comparison, single quantum dot [\onlinecite{Xu2007}] and natural atoms [\onlinecite{Wu1977}] exhibit much lower signal amplifications: 0.005\% and 0.4\%, respectively. Our investigation could address pertinent physics issues in WQED, including dynamics in atom-like mirrors [\onlinecite{Mirhosseini2019}], collective Lamb shift [\onlinecite{Wen2019}], the dynamical Casimir effect [\onlinecite{Wilson2011}], cross-Kerr effect [\onlinecite{Hoi2013a}], generation of non-classical microwaves [\onlinecite{Hoi2012}], probabilistic motional averaging [\onlinecite{Karpov2019}], photon routing [\onlinecite{Hoi2011}], and more.

Driven quantum systems find description through Landau-Zener-St\"{u}ckelberg-Majorana (LZSM) transitions [\onlinecite{Ivakhnenko2022, Shevchenko, Liul2023_2}]. When driven periodically, these systems exhibit interference. LZSM interferometry, crucial for studying fundamental quantum phenomena and characterizing quantum systems, was explored in Refs.~[\onlinecite{Ivakhnenko2022}]. Additionally, quantum logic gates can be implemented using LZSM dynamics [\onlinecite{Campbell2020}].   

This research is an extension of Ref.~[\onlinecite{Liul2023_1}], where the developed theory was approved experimentally. In this article we build the dependencies which were not considered before. The interferograms and dynamic plots presented in this paper allow to understand better underlying physical processes and thus could be interesting and useful for setting up future experiments. Also we solve the Lindblad equation with Lindblad superoperators in different bases: a charge and an energy one and then conclude that the superoperators have to be in the charge basis, since the equation is also written in this basis. Such a comparison is very important from the methodological point of view.     

The rest of the paper is organized as follows. Section~II describes the theoretical aspects of the research: the system Hamiltonian and the Lindblad master equation, which was used for obtaining the results, are introduced. In Sec.~III we study the relations between charge and energy bases. Section~IV shows the obtained interferograms and Sec.~V time dependencies. In Sec.~VI our conclusions are presented.     

\section{Theoretical aspects}
\begin{figure}
	\includegraphics[width=0.95 \linewidth]{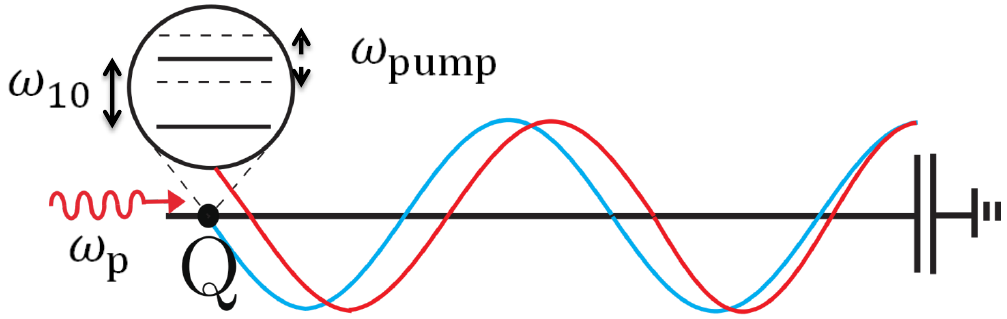}
	\caption{Conceptual sketch of the device. A two-level atom, point-like object (denoted by Q), is coupled to a semi-infinite transmission line waveguide. The atom is located away from the mirror (capacitance). A pump tone with frequency $\omega _{\mathrm{pump}}$ is applied to modulate the transition frequency of the two-level atom. A weak probe tone with frequency $\omega _{\mathrm{p}}$ is applied to the atom-mirror system to measure the reflection coefficient.   
		\label{fig_device}}
\end{figure}

In the study~[\onlinecite{Wen2020}], the experimentally studied reflection coefficient $\abs{r}$ is linked to the theoretically calculated probability of an upper charge level occupation $P_{1}$. An increase in $P_{1}$ corresponds to a decrease in $\abs{r}$. In that article the computations were performed in the charge basis. In our analysis, we maintain this alignment between theoretical predictions and experimental results, conducting calculations in the charge basis. The transmon-type qubit which placed in front of a mirror can be characterized by the Hamiltonian:%
\begin{equation}
H=- \frac{B_{z}}{2}\sigma _{z}-\frac{B_{x}}{2}\sigma _{x},
\label{Hamiltonian}
\end{equation}%
where the diagonal part describes the energy-level modulation 
\begin{equation}
B_{z}/\hbar =\omega _{10}+\delta \sin \omega _{\mathrm{pump}}t,  \label{Bz}
\end{equation}%
the off-diagonal part corresponds to the coupling to the probe signal%
\begin{equation}
B_{x}/\hbar =G\sin \omega _{\mathrm{p}}t.  \label{Bx}
\end{equation}%
To exclude the fast driving from the considered Hamiltonian the authors of Ref.~[\onlinecite{Wen2020}] considered the unitary transformation $U=\exp \left( -i\omega _{\mathrm{p}}\sigma _{z}t/2\right)$
and the rotating-wave approximation [\onlinecite{Silveri15}] and as a result one obtained the new Hamiltonian
\begin{equation}
H_{1}=-\frac{\hbar \widetilde{\Delta \omega }}{2}\sigma _{z}+\frac{\hbar G%
}{2}\sigma _{x},  
\label{H_in_RWA}
\end{equation}%
where%
\begin{eqnarray}
\widetilde{\Delta \omega } &=&\Delta \omega +f(t), \\
\Delta \omega  &=&\omega _{\mathrm{p}}-\omega _{\mathrm{10}}, \\
f(t) &=&\delta \sin \omega _{\mathrm{pump}}t.
\end{eqnarray}%
Here $\delta$ is the energy-level modulation amplitude, $G$ describes the coupling to the probe
signal (Rabi frequency of the probe signal). According to Ref.~[\onlinecite{Wen2020}]
\begin{eqnarray}
G=\frac{\omega_{\mathrm{p}} - \omega_{\mathrm{node}}}{\omega_{\mathrm{node}}}G_{0},
\label{G_expression}
\end{eqnarray}
where $\omega_{\mathrm{node}}$ characterizes the qubit location in a semi-infinite transmission line [corresponding to the blue curve in Fig.~\ref{fig_device}], and $G_{0}$ is proportional to the probe signal amplitude. Moreover, we should mention that if $\omega_{\mathrm{p}} = \omega_{\mathrm{node}}$ the qubit is \textquotedblleft hidden\textquotedblright\ or
\textquotedblleft decoupled\textquotedblright\ from the transmission line, with $G=0$.

To describe the qubit dynamics, we utilize the Lindblad equation, which in the charge basis with the Hamiltonian~(\ref{H_in_RWA}) has the form:%
\begin{eqnarray}
\frac{d\rho}{d t} = -\frac{i}{\hbar}\left [ \widehat{H}_{1}, \rho \right ] + \sum_{\alpha}\breve{L}_{\alpha}\left [ \rho \right ],
\label{Bloch_eq}
\end{eqnarray}
where $\rho~=~\begin{pmatrix} \rho_{00} & \rho_{01} \\ \rho^{*}_{01} & 1 - \rho_{00} \end{pmatrix}$ is the density matrix, such that $P_{1} = 1 - \rho_{00}$. The Lindblad superoperator $\breve{L}_{\alpha}$ describes the system relaxation induced by interactions with the environment,
\begin{eqnarray}
\breve{L}_{\alpha}\left [ \rho \right ] = L_{\alpha}\rho L_{\alpha}^{+} - \frac{1}{2}\left\{L_{\alpha}^{+}L_{\alpha}, \rho \right\},
\label{Lindblad_superoperator}
\end{eqnarray}
where $\left\{a,b \right\} = ab + ba$ is the anticommutator. For a qubit there exist two relaxation channels: dephasing (described by $L_{\phi}$) and energy relaxation (described by $L_{\mathrm{relax}}$). In the energy basis the corresponding operators can be expressed in the following form:%
\begin{eqnarray} 
L_{\mathrm{relax}}^{\mathrm{energy}} = \sqrt{\mathrm{\Gamma}_1}\sigma^{+}, ~ ~
L_{\phi}^{\mathrm{energy}} = \sqrt{\frac{\mathrm{\Gamma}_{\phi}}{2}}\sigma_{z},
\label{L_op_energy}
\end{eqnarray}
with $\sigma^{+} = \begin{pmatrix} 0 & 1 \\ 0 & 0 \end{pmatrix}$, $\sigma_{z} = \begin{pmatrix} 1 & 0 \\ 0 & -1 \end{pmatrix}$, $\mathrm{\Gamma}_1$ being the qubit relaxation, $\mathrm{\Gamma}_2 = \mathrm{\Gamma}_1/2 + \mathrm{\Gamma}_{\phi}$ is the  decoherence rate, $\mathrm{\Gamma}_{\phi}$ is the pure dephasing rate.

\section{Energy and charge basis}
In this section we will discuss the question related to the selection of an appropriate basis. As it was mentioned, the Hamiltonian~(\ref{H_in_RWA}) and the Lindnlad equation~(\ref{Bloch_eq}) were written in a charge basis, while the relaxations $\Gamma_{1}$ and $\Gamma_{\phi}$ are determined in the energy basis. So one should transfer Lindblad operators from the energy basis to the charge one. To study this question let's rewrite the Hamiltonian (\ref{H_in_RWA}) in the form:
\begin{eqnarray}
H(t)=H_0 + V(t),
\end{eqnarray}
where the constant component is equal to 
\begin{eqnarray}
H_0 = \frac{\hbar G}{2} \sigma_x - \frac{\hbar}{2}\left (\omega _{\mathrm{p}}-\omega _{\mathrm{10}} \right) \sigma_z = \frac{\hbar G}{2} \sigma_x - \frac{\hbar}{2}\Delta \omega \sigma_z,
\label{H_0}
\end{eqnarray}
and time-dependent part can be written as
\begin{eqnarray}
V(t) = - \frac{\hbar}{2}\delta \sin \omega _{\mathrm{pump}}t \sigma_z.
\end{eqnarray}
In other words the Hamiltonian (\ref{H_in_RWA}) can be considered as an effective Hamiltonian of a two-level system with tunneling amplitude $G$, and excitation $h(t)~=~-~\frac{\hbar}{2}\left (\Delta \omega + \delta \sin \omega _{\mathrm{pump}}t \right)$.

The transfer matrix which provides transfer from a charge basis to an energy one has a form:
\begin{eqnarray}
\begin{gathered}
			S = \begin{pmatrix} \gamma_+ & \gamma_- \\ \gamma_- & -\gamma_+   \end{pmatrix},
		\end{gathered}
\label{s_matrix_energy}
\end{eqnarray}
where
\begin{eqnarray}
\gamma_{\pm}(t) = \dfrac{1}{\sqrt{2}}\sqrt{1 \pm \frac{\Delta \omega}{\sqrt{\Delta \omega^{2} + G^{2}}}}.
\end{eqnarray}
The matrix $S$ diagonalizes time-independent part of the Hamiltonian $H_0$~(\ref{H_0}): 
\begin{eqnarray}
S^{\dagger} H_0 S = H'_0 = \begin{pmatrix} E_0 &0 \\ 0 & E_1  \\ \end{pmatrix}, 
\end{eqnarray}
where $E_0$, $E_1$~--- are eigenvalues of the Hamiltonian $H_0$.

Therefore, one can transfer Lindblad superoperators $L_{\alpha}^{\mathrm{energy}}$ from energy basis to the charge one $L_{\alpha}$ by using the relation:
\begin{eqnarray}
L_{\alpha} = S L_{\alpha}^{\mathrm{energy}} S^{\dagger}.
\label{L_op}
\end{eqnarray}
Substitution of Eqs.~(\ref{L_op_energy}, \ref{L_op}) to the Lindblad equation (\ref{Bloch_eq}) will give the necessary dependence $P_{1}~=~P_{1}(t)$, which is used for building theoretical plots.

\begin{figure*}
\center{
\includegraphics[width=0.8 \linewidth]{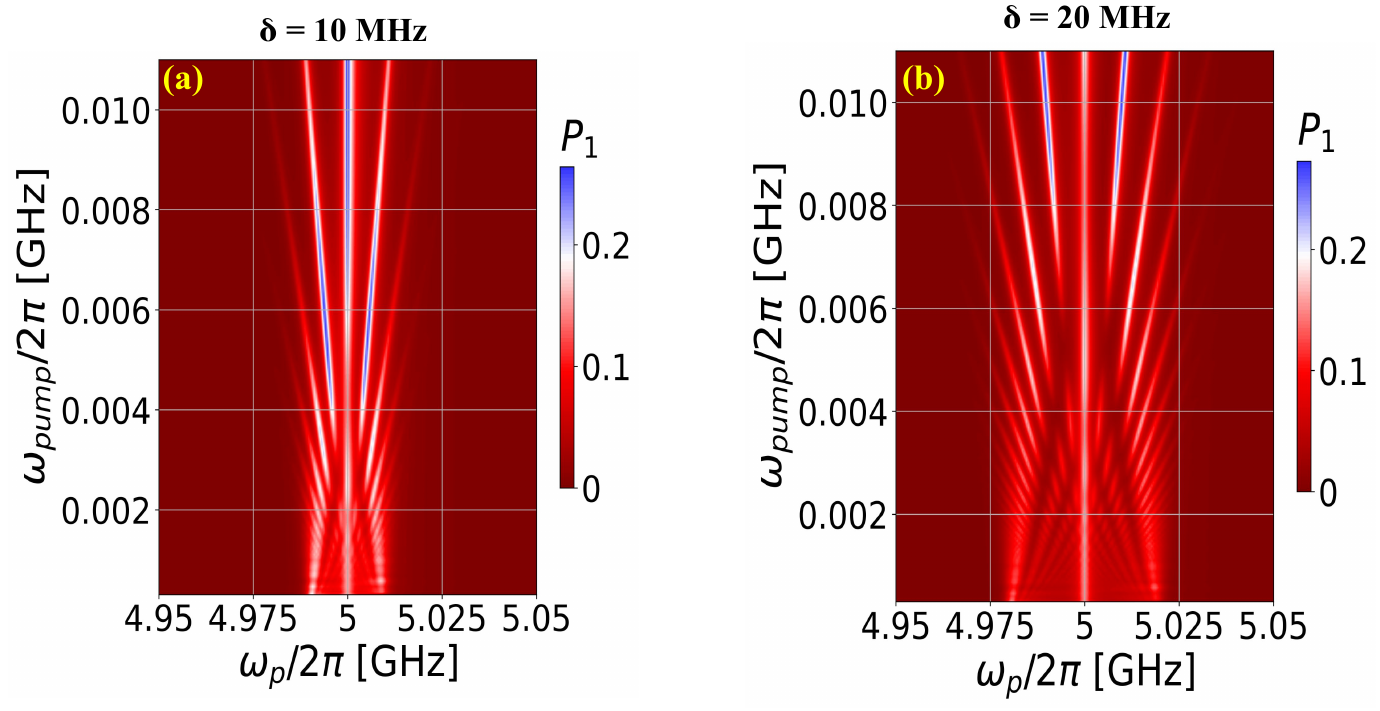}
\caption{LZSM interferograms. The upper charge level occupation probability $P_{1}$ as a function of the pump frequency $\omega_{\rm pump}$ and the probe frequency $\omega_{\rm p}$ at a fixed pump amplitude $\delta$ for $G(\omega_{\rm p}/2\pi~=~5~\mathrm{GHz})~=~2\pi\times0.7~\mathrm{MHz}$. (a)~corresponds to the pump amplitude value $\delta~=~\unit[10]{MHz}$, for (b)~the pump amplitude value is $\delta~=~\unit[20]{MHz}$.
\label{io_freq_interferogram}}}
\end{figure*}

\section{Qubit interferometry}

By solving Eq.~(\ref{Bloch_eq}), we get $P_{1}$ as a function of time $t$, the pump frequency $\omega_{\rm pump}$, the pump amplitude $\delta$, the probe frequency $\omega_{\mathrm{p}}$, the signal amplitude $G$. The occupation probability of the upper charge level is a function of all these parameters, $P_{1}~=~P_{1}(t,~ \omega_{\rm pump},~ \omega_{\rm p},~ \delta,~ G)$. We can also calculate the dependencies for $P_{1}$ in the stationary regime by averaging the results over time.

To obtain the time-averaged values, we analyzed the curve $P_{1}~=~P_{1}(t)$ to select the minimum time $t_{\mathrm{min}}$ after which the amplitude of oscillations does not change. And then averaging was applied for the interval $[t_{\mathrm{min}}, t_{\mathrm{final}}]$, where $t_{\mathrm{final}}$ corresponds to the pulse off time. We determined that for our case $t_{\mathrm{min}}~=~1.5~\mu {\rm s}$ and $t_{\mathrm{final}}~=~2.0~\mu {\rm s }$ (see Ref.~[\onlinecite{Liul2023_1}] for more details).

In this section, we theoretically investigate the dependence of the qubit upper charge level occupation probability $P_{1}$ on the pump frequency $\omega_{\rm pump}$ and the probe frequency $\omega_{\rm p}$. Note that the experimental study of such dependence was not conducted. But since one found the values of the fitting parameters in Ref.~[\onlinecite{Liul2023_1}] we can construct such a dependence. It is shown in Fig.~\ref{io_freq_interferogram}, (a)~corresponds to the pump amplitude $\delta~=~\unit[10]{MHz}$, for (b)~the pump amplitude is equal to $\delta~=~\unit[20]{MHz}$. From the analysis of interferograms, it becomes clear that the larger the pumping amplitude $\delta$, the larger the amplitude along the $\omega_{\rm p}$ axis. Similar dependencies can be observed in Ref.~[\onlinecite{Wen2020}], where only the stationary case was considered.

\begin{figure*}
\center{
\includegraphics[width=0.95 \linewidth]{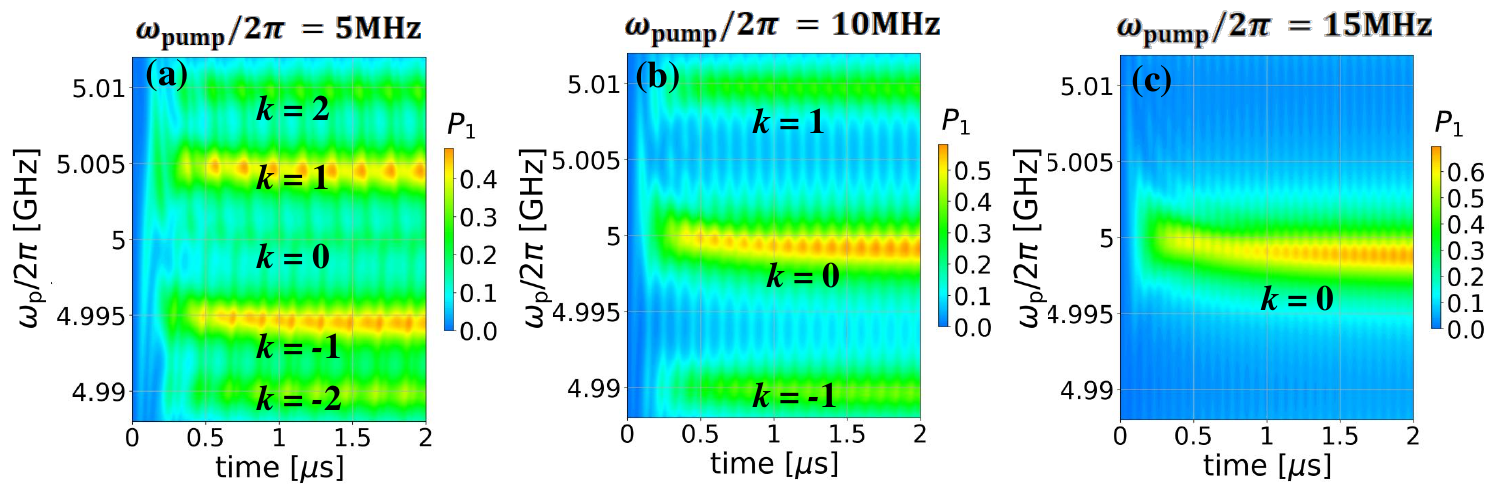}
\caption{Calculated dynamics of a transmon-type qubit with Lindblad operators in the energy basis. Dependence of the upper charge level occupation probability $P_{1}$ on time $t$ and the probe frequency $\omega_{\rm p}$ at different pump frequencies $\omega_{\rm pump}$. During the calculations, the following values were used: the probe amplitude $G(\omega_{\rm p}/2\pi~=~5~\mathrm{GHz}) = 2\pi\times1.4~\mathrm{MHz}$, the pump amplitude $\delta~=~\unit[10]{MHz}$. (a, b, c) show plots constructed from data calculated theoretically for the case of Lindblad operators in the energy basis. The qubit is irradiated with a pump frequency of (a)~$\omega_{\rm pump}/2\pi~=~\unit[5]{MHz}$, (b)~$\omega_{\rm pump}/2\pi~=~\unit [10]{MHz}$, (c)~$\omega_{\rm pump}/2\pi~=~\unit[15]{MHz}$.
\label{Fig_time_f_pump_energy}}}
\end{figure*}
\begin{figure*}
\center{
\includegraphics[width=0.95 \linewidth]{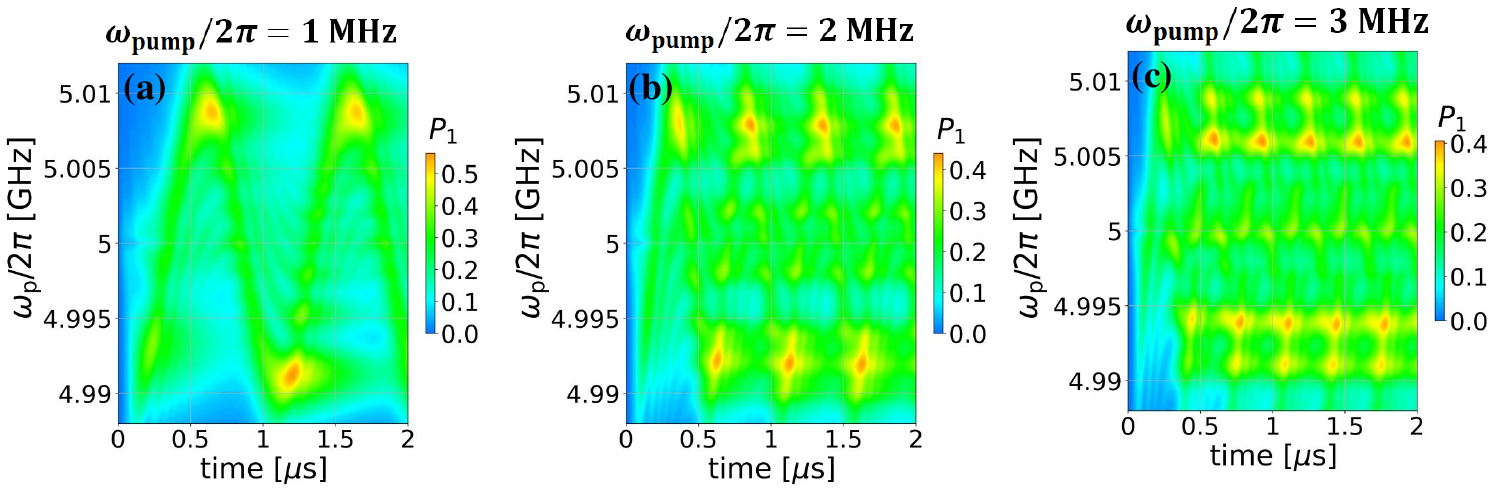}
\caption{Coherent dynamics of a transmon-type qubit at low frequencies of the $\omega_{\rm pump}$ pump signal. The upper charge level occupation probability $P_{1}$ as a function of time $t$ and the probe frequency $\omega_{\rm p}$ at different pump frequencies $\omega_{\rm pump}$. The following parameter values were used for the calculations: amplitude of the probe signal $G(\omega_{\rm p}/2\pi=5~\mathrm{GHz})~=~2\pi\times1.4~\mathrm{MHz}$, amplitude of the pump signal $\delta~=~\unit[10] {MHz}$. The qubit is irradiated by pump signal with a frequency of (a)~$\omega_{\rm pump}/2\pi~=~\unit[1]{MHz}$, (b)~$\omega_{\rm pump}/2\pi~=~\unit [2]{MHz}$, (c)~$\omega_{\rm pump}/2\pi~=~\unit[3]{MHz}$.
\label{io_time_interferogram_low_freq}}}
\end{figure*}

Such interferograms are useful not only for obtaining fitting parameters between theory and experiment, but they also play a key role in system characterization. In particular, they can help to:
\begin{itemize}
      \item estimate the system decoherence time;
      \item provide a tool for calibrating signal strength during experimental studies;
      \item open up new possibilities for multiphoton spectroscopy.
\end{itemize}

\section{Qubit dynamics}

\subsection{Dependence of the upper charge level occupation probability on the probe frequency for the case of Lindblad operators in the energy basis}

\begin{figure*}
\center{
\includegraphics[width=0.95 \linewidth]{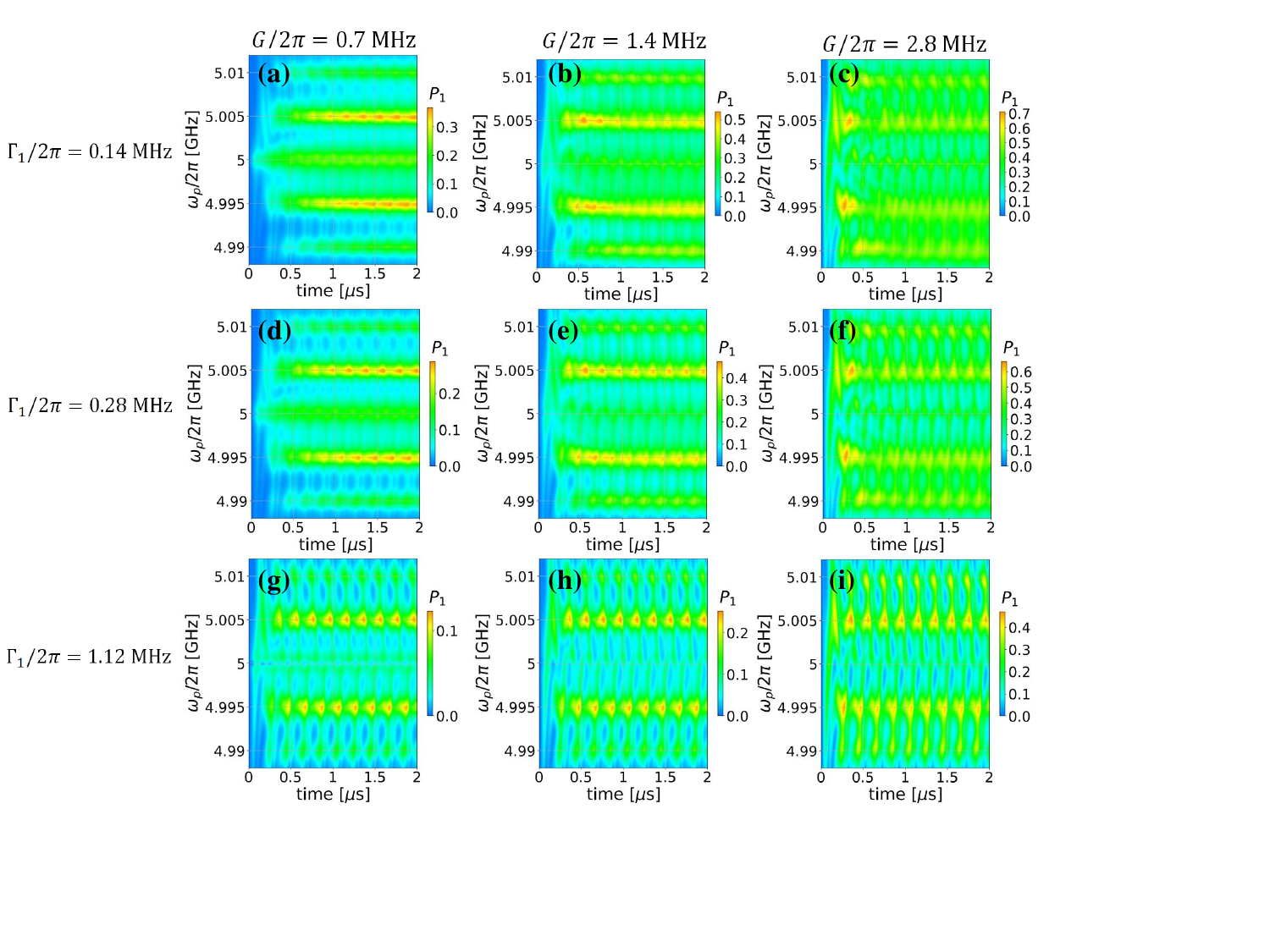}
     \caption{Dependence of the upper charge level occupation probability on time $t$ and the probe frequency $\omega_{\rm p}$ at different values of the probe amplitude $G$ and the relaxation rate $\Gamma_{1}$ for the pump frequency $\omega_{\rm pump}/2\pi~=~\unit[5]{MHz}$. For calculations, the value of the pump amplitude was taken equal to $\delta~=~\unit[10]{MHz}$. The values of the probe amplitude $G$ change from left to right and have the values $G/2\pi~=~\unit[0.7]{MHz}$, $G/2\pi~=~\unit[1.4]{MHz}$, $G /2\pi~=~\unit[2.8]{MHz}$. The relaxation rate changes from top to bottom and has the value $\Gamma_{1}~=~\unit[0.14]{MHz}$, $\Gamma_{1}~=~\unit[0.28]{MHz}$, $\Gamma_{1}~=~\unit[1.12]{MHz}$ respectively.
\label{io_G_Gamma_time_interf}}}
\end{figure*}

This section presents the results of calculations for the case of Lindblad operators in the energy basis. Note, that this approach is not correct and the obtained theoretical dependencies are not consistent with the experimental ones, however, from a methodological point of view, it is useful to give such an example. Figure~\ref{Fig_time_f_pump_energy} shows the pictures obtained as a result of such calculations. This dependence corresponds to Fig.~4(d-f) in Ref.~[\onlinecite{Liul2023_1}].

From the comparison of the theoretical (Fig.~\ref{Fig_time_f_pump_energy}(a-c)) and the corresponding experimental (Fig.~4(a-c) in Ref.~[\onlinecite{Liul2023_1}]) pictures, we can conclude that the Lindblad operators have to be transferred to the charge basis. The obtained dependencies have a behavior similar to the experimental ones, for example, resonances occur at $\Delta \omega=k\omega_{\rm pump}$ both in the theory and in the experiment. However, unlike the experimental ones in the theoretical pictures, the resonance $k=0$ is slightly shifted from the line $\omega_{\rm p}/2\pi = \unit[5]{GHz}$ and it is more blurred (wider).

\subsection{Dependence of the upper charge level occupation probability on the probe frequency and time}
In the presented section, we will construct dependencies similar to those shown in the work~[\onlinecite{Liul2023_1}] in Fig.~4(d,e,f), but at lower values of the frequency of the pump signal $\omega_{\rm pump}$. Figure~\ref{io_time_interferogram_low_freq} shows the obtained interferograms. The qubit is irradiated by pump signal with a frequency of (a)~$\omega_{\rm pump}/2\pi~=~\unit[1]{MHz}$, (b)~$\omega_{\rm pump}/2\pi ~=~\unit [2]{MHz}$, (c)~$\omega_{\rm pump}/2\pi~=~\unit[3]{MHz}$.

In Fig.~\ref{io_time_interferogram_low_freq}(a), for the case $\omega_{\rm pump}/2\pi~=~\unit[1]{MHz}$, the resonance peaks are not distinguished, this may occur due to the fact that they are located too close to each other and therefore merge. For the Fig.~\ref{io_time_interferogram_low_freq}(b), strong peaks are observed only at a distance $\omega/2\pi~=~\unit[8]{MHz}$ from the line $\omega_{\rm p}/2 \pi~=~\unit[5]{GHz}$, for the Fig.~\ref{io_time_interferogram_low_freq}(c) strong peaks are observed only at a distance $\omega/2\pi~=~\unit[6]{MHz},~\unit[9]{MHz}$ from the line $\omega_{\rm p}/2 \pi~=~\unit[5]{GHz}$, while peaks at a distance of $\omega/2\pi~=~\unit[3]{MHz}$ are weakly highlighted.

\subsection{Dependence of the upper charge level occupation probability on the probe amplitude and the probe frequency for at different values of the probe amplitude and the relaxation rate}

In the presented section, we will investigate the dependence of the upper charge level occupation probability on time $t$ and the probe frequency $\omega_{\rm p}$ at different values of the probe amplitude $G$ and the relaxation rate $\Gamma_{1}$ for the pump frequency $\omega_{\rm pump}/2\pi~=~\unit[5]{MHz}$. For calculations, the pump amplitude is taken equal to $\delta~=~\unit[10]{MHz}$. The resulting dependencies are shown in Fig.~\ref{io_G_Gamma_time_interf}.

Analyzing the obtained dependencies, it can be concluded that with an increase of the relaxation rate $\Gamma_{1}$ at a fixed value of the probe amplitude $G$ (within a certain column we move from top to bottom), the resonances become less blurred, that is, their amplitude along the time axis $ t$ decreases. This fact is understandable, because with an increase in the relaxation rate, the system quickly relaxes to the unexcited state, what is observed on the obtained dependencies.

For the case $G/2\pi~=~\unit[2.8]{MHz}$, the first resonances observed at $t$ $\approx$ 0.25~$\mu {\rm s}$ are clearly expressed, while for probe amplitude $G/2\pi~=~\unit[1.4]{MHz}$, the maximum value of the upper charge level occupation probability $P_{1}$ corresponds to the second and third resonances, while the first resonance is weakly highlighted. For $G/2\pi~=~\unit[0.7]{MHz}$ and $\Gamma_{1}~=~\unit[0.14]{MHz}$, $\Gamma_{1}~=~\unit[0.28]{MHz }$ during the first $t$~$\approx$~1~$\mu {\rm s}$ the value of the upper charge level occupation probability $P_{1}$ has smaller values than for longer time values. In the case of $G/2\pi~=~\unit[0.7]{MHz}$ and $\Gamma_{1}~=~\unit[1.12]{MHz}$ the first two resonances ($t$~$<$~0.5~$ \mu {\rm s}$) are highlighted more weakly than the following ones.

\section{Conclusions}
 We considered the dynamics and stationary mode of a transmon-type capacitive shunted qubit in front of a mirror, which is affected by two signals: probing and pumping.

To analyze the steady state, the upper charge level occupation probability $P_1$ was calculated as a function of the pump frequency $\omega_{\rm pump}$ and the probe frequency $\omega_{\rm p}$ at the fixed pump amplitude $\delta$ and the probe amplitude $G$. The resulting dependencies can be used to obtain fitting parameters; system decoherence time estimation; power calibration; studying multiphoton spectroscopy.

To analyze the dynamics, we plotted the dependencies of the upper charge level occupation probability $P_1$ on (a)~time $t$ and the probe frequency $\omega_{\rm p}$ at different pump frequencies $\omega_{\rm pump}$; (b)~time $t$ and the probe frequency $\omega_{\rm p}$ at different values of the probe amplitude $G$ and the relaxation rate $\Gamma_{1}$ for the pump frequency $\omega_{\rm pump }/2\pi~=~\unit[5]{MHz}$. The obtained results can be used in setting up future experiments.   

The Lindblad equation with Lindblad superoperators in different bases: a charge and an energy one was solved. One concluded that the superoperators had to be in the charge basis, since the equation is also written in this basis. Such a comparison should be useful from the methodological point of view.

\begin{acknowledgments}
The author acknowledge fruitful discussions with S.~N.~Shevchenko, A.~I.~Ryzhov, O.~V.~Ivakhnenko. The author was partially supported by the grant from the National Academy of Sciences of Ukraine for research works of young scientists (1/H-2023, 0123U103073).
\end{acknowledgments}

\nocite{apsrev41Control} 
\bibliographystyle{apsrev4-1}
\bibliography{references}

\end{document}